\documentclass[12pt,preprint]{aastex}
\usepackage[dvipsnames]{color} 

\newcommand{\beq}{\begin{equation}}
\newcommand{\eeq}{\end{equation}}
\newcommand{\barray}{\begin{eqnarray}}
\newcommand{\earray}{\end{eqnarray}}
\newcommand{\bfig}{\begin{figure}}
\newcommand{\efig}{\end{figure}}
\newcommand{\cs}{c_{\rm s}}
\newcommand{\csq}{c_{\rm s}^2}

\newcommand{\msun}{M_{\sun}}

\slugcomment{To appear in The Astrophysical Journal}
\shorttitle{Brown Dwarf and Low Mass Star Ejection}
\shortauthors{Shantanu Basu and E. I. Vorobyov}
\begin{document}

\title{A Hybrid Scenario for the Formation of Brown Dwarfs and Very Low Mass Stars}

\author{Shantanu Basu\altaffilmark{1}, Eduard I. Vorobyov\altaffilmark{2,}\altaffilmark{3} }
\altaffiltext{1}{Department of Physics and Astronomy, University of Western Ontario,
London, Ontario, N6A 3K7, Canada; basu@uwo.ca.}
\altaffiltext{2}{Institute of Astronomy, The University of Vienna, Vienna, 1180, Austria; 
eduard.vorobiev@univie.ac.at.} 
\altaffiltext{3}{Institute of Physics, Southern Federal University, Stachki 194, 
Rostov-on-Don, 344090, Russia.}

\begin{abstract}{}
We present a calculation of protostellar disk formation and evolution in which
gaseous clumps (essentially, the first Larson cores formed via disk fragmentation) 
are ejected from the disk during the early stage of 
evolution. This is a universal process related to the phenomenon of ejection
in multiple systems of point masses. However, it occurs in our model entirely 
due to the interaction of compact, gravitationally-bound gaseous clumps and
is free from the smoothing-length
uncertainty that is characteristic of models using sink particles. 
Clumps that survive ejection span a mass 
range of 0.08--0.35 $M_\odot$, and have ejection velocities 
$0.8 \pm 0.35$ km s$^{-1}$, which are several times greater than the escape speed.
We suggest that, upon contraction, 
these clumps can form substellar or low-mass stellar objects with notable disks, 
or even close-separation very-low-mass binaries. 
In this hybrid scenario, allowing for ejection of clumps rather than 
finished protostars/proto--brown-dwarfs, disk formation and the low velocity dispersion of low-mass
objects are naturally explained, while it is also consistent with the observation of isolated
low-mass clumps that are ejection products.
We conclude that clump ejection and the formation of isolated
low mass stellar and substellar objects is a common occurrence, with important 
implications for understanding the initial mass function, the brown dwarf desert, 
and the formation of stars in all environments and epochs.
\end{abstract}

\keywords{accretion, accretion disks --- hydrodynamics --- instabilities
--- ISM: clouds ---  planets and satellites: formation --- stars: formation -- stars: low mass, brown dwarfs}

\section{Introduction}

The formation of brown dwarfs (BDs) is usually attributed to either of
two distinct mechanisms.
One model is the direct collapse of a very low mass cloud fragment
whose mass straddles the substellar mass limit. 
Formation mechanisms for such low mass (and consequently high density)
cores include colliding flows in a turbulent magnetic medium \citep{pad04} 
and photoerosion of more massive cores 
by the ionizing UV radiation from nearby OB stars \citep{Whitworth04}.
Since the collapse of a
prestellar core has been shown by numerous studies to lead to a very focused
collapse with a power-law density profile, the probability of fragmentation during
the runaway collapse phase (i.e., before the formation of a central hydrostatic object)
is low. This leads to the scenario that
the stellar and substellar mass function is simply a scaled-down version of
the core mass function. 

However, during the protostellar phase
(i.e., after the formation of a central hydrostatic object), 
and unlike in the prestellar phase,
the relative scaling of flow variables in the presence of angular momentum
can lead to the
formation of a centrifugally-balanced disk \citep{bas95a}.  
Such a disk has a high density and evolves on a time scale much longer than 
its dynamical time.
Therefore, the protostellar disk has time for fragmentation if conditions 
are suitable, e.g., the criteria for gravitational instability and fragmentation are
satisfied \citep{too81,Gammie01}. Furthermore, fragmentation may be induced
by close encounters with members of a stellar cluster in otherwise gravitationally stable  
disks \citep[e.g.,][]{Thies10}. This leads to the second model: that 
disk fragmentation is responsible for
the formation of low mass and substellar objects in the vicinity of a forming star,
even if the mass of the parent core is far greater than the substellar limit.
In this scenario, free-floating BDs would be the result of ejections from the disk, which is
a plausible outcome based on extrapolation from three-body calculations 
of interacting point masses and can account for peculiar properties of BDs as compared to those
of low-mass stars \citep{Kroupa03}. \citet{bat09} obtained the BD ejection 
from SPH modeling of molecular cloud fragmentation 
resolved from pc scales down to the opacity limit that occurs on 
AU scales. In that model, multiple collapsing fragments each form a disk. Some
disks undergo fragmentation, and low-mass objects, treated as point masses
in the SPH approximation, can be ejected from multiple systems.
In the models of \citet{bat09}, a sink cell is introduced on AU scales and gas
is cleared away from a region around them of size up to 10 AU.
Furthermore, \citet{sta09} have modeled the formation and ejection of 
BDs after disk fragmentation, also using SPH and sink cells, and 
starting from an initial condition of a pre-existing massive disk 
that has a mass equal to that of the central star.

In this paper, we pursue models that solve the self-consistent formation of 
a centrifugal disk from the collapse of a larger cloud core, employing the 
thin-disk approximation. Computational efficiency allows us to run a large 
number of models, with various initial conditions of cloud mass and angular
momentum, as well as study the long-term evolution of disks. 
We employ a central sink cell of size 6 AU, representing a central star and some 
circumstellar material, but the surrounding disk
that forms is followed without recourse to sinks, even when fragmentation occurs.
Our main result is that ejection of protostellar/protobrown dwarf mass {\it clumps} occurs
through the interaction of multiple gravitationally-bound gaseous clumps,  
which are essentially the first Larson cores formed via gravitational fragmentation 
of protostellar disks. 

Our results support a new hybrid paradigm of BD and low mass star formation by 
disk fragmentation followed by clump ejection (rather than by ejection of 
finished low-mass stars/brown dwarfs). This scenario naturally accounts
for the presence of disks around BDs and the relatively low ejection speeds
in this scenario can account for the observed velocity dispersion and
physical location of BDs relative to YSOs. It is also consistent with 
the so-called brown dwarf desert, i.e., the  
lack of BD companions to primary stars at distances less than several tens of AU
\citep{mcc04,mar05}. A disk fragmentation origin for BDs can 
explain the brown dwarf desert \citep[see, e.g.,][]{sta11} since 
cooling constraints in disks
\citep{raf07} favor fragmentation at radii 
$\gtrsim 50$ AU from the primary object. Furthermore, fragments that do
form at closer range or are torqued inward from wider orbits are often 
subject to a runaway inward migration toward merger with the central object \citep{vor05,vor06}, so that low mass gaseous companions are not 
typically expected to settle into stable inner orbits. However, it has been 
suggested that some of the eventually tidally disrupted clumps may have 
enough time to build up solid cores. These may then be released into inner 
orbits after the disruption, providing thermally-processed solids or even
rocky planetary cores \citep{bol10,Nayakshin10,Cha11,vor11}. This
scenario may account for planet formation in the inner disk but presumably 
not something as massive as a BD. Detailed simulations of this
possible mechanism remain to be worked out. 

In contrast, the direct core collapse paradigm for BD formation also suffers from the
need to explain very low mass gravitationally bound objects that are one to two
orders of magnitude less than the ambient Jeans mass in molecular clouds.
This can be accomplished with very high bounding pressure, whether
due to very strong magnetic fields \citep{bas95b} or high turbulent
pressure \citep{pad02,cha11}. Both the ideas of extremely high magnetic or
turbulent bounding pressure on a fragment suffer from a lack of direct
observational evidence at the moment \citep[see][]{and09}. 
Turbulent pressure may also not be present in such a focused manner as
to act like an isotropic bounding pressure.

\section{Model}
\label{model}

Our model and method of solution are the ones presented in detail by
\citet{vor10}, which is an update to the model presented by
\citet{vor06}. We solve the mass, momentum, and energy transport 
equations in the thin-disk approximation, using vertically-integrated
quantities. Disk self-gravity is calculated using a two-dimensional
Fourier convolution theorem for polar coordinates. The energy equation
includes compressional heating, heating due to stellar and background irradiation, radiative
cooling from the disk surfaces, heating due to turbulent viscosity (parameterized 
via the usual Shakura \& Sunyaev prescription with $\alpha=0.005$), 
and a diffusion approximation to link
the effective surface temperature of the disk with the midplane 
temperature. Frequency-integrated opacities are adopted from the 
calculations of \citet{bel94} and a smooth transition is 
introduced between the optically thin and optically thick regimes.
In the opaque regime, gravitationally-bound clumps that form in 
the disk are essentially ``first hydrostatic cores'' or ``first Larson cores'' 
in the parlance of star formation \citep{lar69,mas00}. A ``second collapse''
down to stellar dimensions can take place when the central temperature
reaches $\sim 2000$ K and H$_2$ is dissociated. This possibility is not 
allowed in our energy equation so that we may continue to track the 
evolution without recourse to sink cells\footnote{Resolving second cores with a typical size of
several stellar radii, and thus avoiding the use of sink particles, is a formidable task for grid-based codes designed to study the 
{\it global} evolution of protostellar disks as it requires a local spatial 
resolution $\la 0.01$~AU.}, however clumps in the disk 
typically do not reach this temperature before disruption or ejection
(see Section 4).  
The stellar irradiation is based on a luminosity that is a
combination of accretion luminosity and photospheric luminosity
calculated from the pre-main-sequence tracks of \citet{dan97}.
The pressure is related to the internal energy through an ideal gas
equation of state.

Models presented in this paper are run on a polar coordinate ($r,\phi$)
grid with $512 \times 512$ zones. A central sink cell of radius 6 AU
is employed and the potential of a central point mass is added to the 
disk self-gravity once a central hydrostatic core is formed. 
The radial points are logarithmically
spaced, with the innermost cell outside the central sink having size 0.07--0.1
AU depending on the cloud core size (i.e., the radius
of the computational region). The latter varies in the 0.025--0.12~pc (5000--24000~AU) 
limits. The radial and azimuthal resolution are about 1 AU 
at a radius of 100 AU. 
The inner and outer boundary conditions are set to allow for free 
outflow from the computational domain.
The initial conditions for the prestellar core correspond to profiles
of column density $\Sigma$ and angular velocity $\Omega$ of the form
\barray
\Sigma & = & {r_0 \Sigma_0 \over \sqrt{r^2+r_0^2}}\:, \\
\Omega & = &2\Omega_0 \left( {r_0\over r}\right)^2 \left[\sqrt{1+\left({r\over r_0}\right)^2
} -1\right].
\label{ic}
\earray
These profiles have a small near-uniform
central region of size $r_0$ and then transition to an $r^{-1}$ profile;
they are representative of a wide class of observations and theoretical models
\citep{bas97,and09,dap09}. Our previous papers have also shown that 
the qualitative features of the ultimate centrifugal disk evolution are
independent of the specific profiles adopted for the initial prestellar core.
Our reference model presented in this paper has rotational support
in the initial state characterized by the dimensionless parameter
$\eta \equiv \Omega_0^2r_0^2/\csq = 1.4\times 10^{-2}$, where $\cs = 0.188$ km s$^{-1}$
(corresponding to a temperature $T=10$ K) in the 
initial state but varies spatially and temporally during the evolution. 
Alternately, the ratio of rotational
energy to the magnitude of gravitational potential energy is 
$\beta = 1.3 \times 10^{-2}$. Table 1 shows a list of
parameters and some results from models we have run for this study.
The reference model of our study (model 5) has $r_0 = 2006$~AU, a total core mass 
$M_{\rm c} = 0.92\,\msun$, and $\beta = 1.3 \times 10^{-2}$.

\begin{table*}
\center
\caption{Models showing ejection and properties of ejected clumps}
\label{table1}
\begin{tabular}{cccccccccc}
\hline\hline
Model & $\beta$ & $\Omega_0$ & $r_0$ &  $M_{\rm c}$ & $M_{\rm s}$  & $M_{\rm eject}$  & $t_{\rm eject}$ & $v_{\rm eject}$
& $v_{\rm esc}$ \\
\hline
$1^\dagger$ & $0.23\times 10^{-2}$ & 0.48 & 3940 & 1.78 & 1.03 & 0.13 & 0.52 & 0.38 & 0.34 \\
$2^\star$ & $0.56\times 10^{-2}$ & 1.14 & 2400 & 1.08  & 0.68 & 0.15 & 0.34 & 0.77 & 0.31 \\
$3^\dagger$ & $0.56\times 10^{-2}$ & 0.72 & 3770 & 1.69 & 0.94 & 0.06 & 0.43 & $0.6^\diamond$ & 0.33 \\
$4^\dagger$ & $0.56\times 10^{-2}$ & 0.88 & 3085 & 1.38 & 0.82 & 0.05 & 0.70 & $0.5^\diamond$ & 0.36 \\
5 (ref.) & $1.3\times 10^{-2}$ & 2.0 & 2006 & 0.92  & 0.47 & 0.15 &   0.30 &   0.91 & 0.31 \\
6 & $1.3\times 10^{-2}$ & 1.2 & 3430 & 1.54 & 0.45/0.69 & 0.2/0.2  & 0.24/0.65 & 0.88/1.20  & 0.26/0.27 \\
7 & $1.3\times 10^{-2}$ & 1.0 & 4115 & 1.85  & 0.86/0.87 & 0.08/0.2 & 0.73/0.83 & 0.76/0.52 & 0.42/0.29 \\
$7^\dagger$ & -- & -- & -- & --  & 0.73 & 0.11 & 0.45 & $0.7^\diamond$ & 0.45 \\
8 & $1.3\times 10^{-2}$ & 3.4 & 1200 & 0.54 & 0.31 & 0.1 & 0.18 & 1.0 & 0.32 \\
$9^\dagger$ & $1.3\times 10^{-2}$ & 1.5 & 2780 & 1.25  & 0.49 & 0.2 & 0.22 & $0.45^\diamond$ & 0.29 \\
$10^\star\dagger$ & $2.24\times 10^{-2}$ & 2.9 & 1885 & 0.85 & 0.46 & 0.08 & 0.54 & 0.44 & 0.28 \\
$11^\star$ & $2.24\times 10^{-2}$ & 1.4 & 3940 & 1.77  & 0.51 & 0.34  & 0.33 & 0.45 & 0.32 \\
$11^\dagger$ & -- & -- & -- & -- & 0.56 & 0.09 & 0.45 & $0.62^\diamond$ & 0.38\\
\hline
\end{tabular}
\tablecomments{All distances are in AU, velocities in km~s$^{-1}$, angular
velocities in km~s$^{-1}$~pc$^{-1}$, time in Myr, and masses in $\msun$.}
\end{table*}

\section{Results}
\label{results}

Our recent numerical hydrodynamics simulations with realistic cooling 
have shown that fragment
formation is enhanced at radii $\gtrsim 50$ AU (in agreement with many previous studies
on disk fragmentation), and that episodic
accretion due to clump formation and inspiralling of the clumps
is a robust result \citep{vor10}. In this paper, we report on a 
qualitatively new kind of result --- an occasional ejection of fragments from the disk 
into the intracluster medium. Long integration times ($\sim 1$~Myr) 
need to be run in order to capture the occasional ejection events.  Below, we discuss in detail
the results for our prototype model.

Figure 1 shows a sequence of column density images for our reference model (model~5),
at various times after the formation of a central object, and in a region
of size 2400 AU on each side. The circumstellar disk forms at $t\approx0.01$~Myr
and already at $t=0.05$~Myr it undergoes fragmentation.
Within the first $\approx 0.25$ Myr, multiple fragments are formed in the 
relatively massive disk, at distances of $\gtrsim 50$ AU and $\lesssim$
few hundred AU. They are generally torqued inward through gravitational
interaction with trailing spiral arms, as first found by \citet{vor05,vor06}.
Others located at large radii may eventually disperse. However,
under sometime favorable conditions, a clump within a multi-clump  environment 
can be ejected through many-body interaction. The ejection is also aided by the 
non-axisymmetric potential of the relatively massive disk.
The arrow in the image in the
second row of the middle column points to a clump that is subsequently ejected from 
the system.  The velocity of the ejected clump (at the moment when it leaves the
computational boundary at $r=12000$~AU) is about 0.9~km~s$^{-1}$,
which is a factor of 3 greater than the escape velocity $v_{\rm esc}=(2 G M_{\rm sd}/r)^{1/2}$,
where the total mass of the central star and its disk is $M_{\rm sd}=0.6~M_\odot$.
This  means that 
the clump will truly be lost to the system. The
ejection event is transient and the clump is not seen in the subsequent image at $t=0.27$~Myr. 
The {\it total} mass of the clump, calculated as the mass passing the computational boundary during
the ejection event, is $0.15~M_\odot$. It should be noted, however, that this value includes
not only the compact core but also a diffuse envelope and even fragments of spiral arms (see Fig.~2
below). 
We speculate that upon contraction this clump may form a substellar object, given 
that a significant fraction of mass remains in the disk until it is ejected due to outflows
and/or dispersed due to photoevaporation.
No further major fragmentation episodes in the protostellar disk are seen after $t=0.27$~Myr,
likely due to a sharp drop in the total disk mass caused by the ejection. The disk then
gradually evolves toward a nearly axisymmetric state. 

To illustrate the ejection mechanism in more detail, 
Figure 2 shows a sequence of column density images
on a much larger spatial scale of the simulation box, 20000 AU on each side, and on a much
shorter time scale, $t=$0.26--0.31~Myr.
The arrows in the images identify the same clump that
was identified in Figure 1, and show its ejection from the disk and indeed the
entire simulation box. As the clump moves nearly radially outward, it grows in size and diffuses
somewhat but this is an artefact of the decreasing resolution of our numerical grid at larger radii.
Future higher resolution simulations will be 
needed to verify the evolution of the clumps after they undergo an 
ejection event. 

\bfig
  \resizebox{\hsize}{!}{\includegraphics{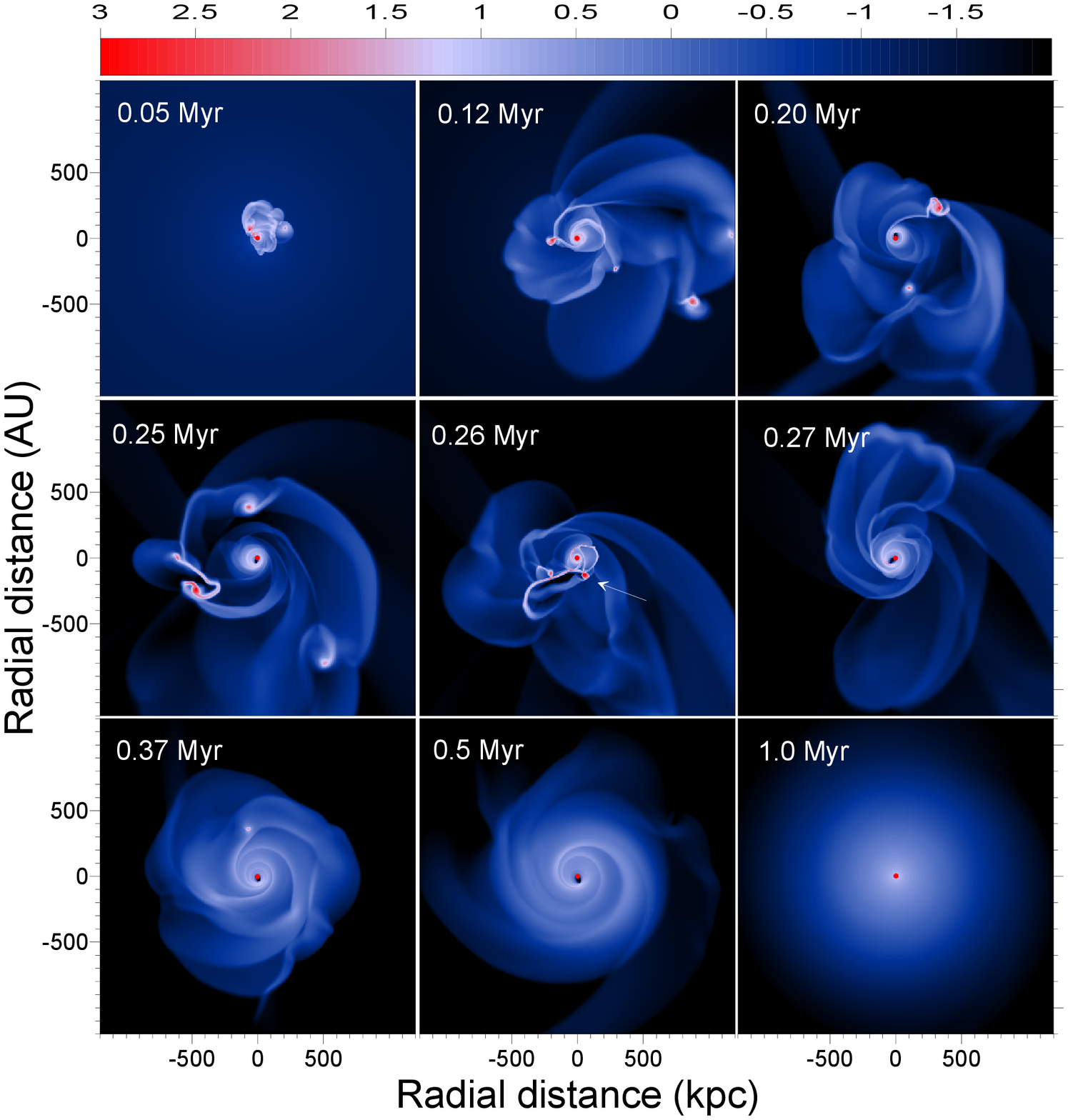}}
      \caption{Gas surface density distribution (g cm$^{-2}$, log units) in the
reference model at several time instances after the formation of a central
star. The box size is 2400 AU on each side, and represents a small subregion of the
overall computational domain. An arrow identifies a clump that is ejected from the
inner region soon afterward.}
         \label{fig1}
\efig

\bfig
  \resizebox{\hsize}{!}{\includegraphics{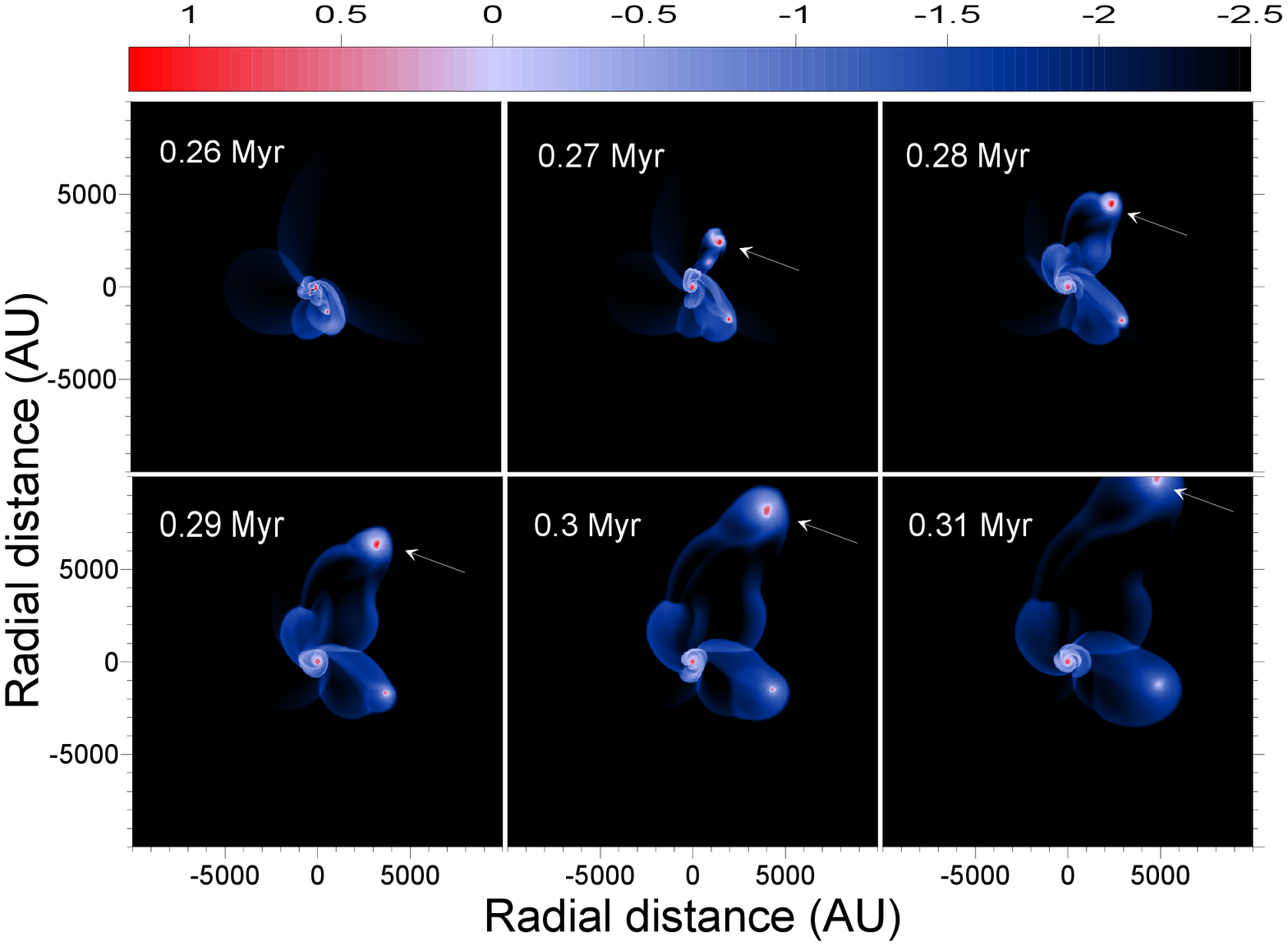}}
      \caption{Gas surface density distribution (g cm$^{-2}$, log units) in the
reference model at several time instances after the formation of a central
star. The box size is 20000 AU on each side, and represents nearly the full extent
of the computational domain. Arrows identify a clump that is ejected from the
system after a multi-body interaction within the centrifugal disk.}
         \label{fig2}
\efig

Figure 3 shows the mass accretion rate onto the central sink as a function of time
since the beginning of numerical simulations. 
The time $t=0.12$~Myr represents the moment that a central object is formed. 
After a brief period of smooth accretion, a disk forms outside the sink cell, and the
subsequent accretion is highly variable, with the highest rates exceeding 
$10^{-4}~M_\odot$~yr$^{-1}$ and the lowest rates plunging below $10^{-9}~M_\odot$~yr$^{-1}$.
Interestingly, there is a 
late burst at 0.38 Myr (0.26~Myr after the formation of the central star) 
that is greater in magnitude than all previous bursts.
This burst corresponds to the inward flow of material that accompanies the 
ejection of the clump. The effects are twinned due to the need for overall angular
momentum conservation---one clump is ejected while the other one is driven into the center.  
After this last major burst, the disk settles down to 
a more quiescent phase of evolution with a much lower accretion variability, characterized
by flickering in the accretion rate but no bursts, and a gradual decline in 
accretion rate. 
The residual accretion rate of a few $\times\, 10^{-8}\, \msun$ yr$^{-1}$
at 1.0~Myr is consistent with observed accretion rates for T Tauri stars
\citep[e.g.,][]{cal04}. This behavior agrees well with a gradual 
gravitational stabilization of the disk that is caused the ejection event and the associated 
substantial mass loss. From this time onward, viscous torques
start to dominate gravitational torques in the disk mass transport
\citep{VB09}.

Figure 4 illustrates the evolution of mass in various components of the system, as 
well as in the overall simulation region. The total mass in the system suffers a 
drop as the ejected clump exits through the outflow outer boundary applied to the simulation
region. The mass loss equals the clump mass: $\approx 0.15\, \msun$. The clump
maintains its form all the way to the edge of our simulation box 
notwithstanding a constantly decreasing resolution on the logarithmic grid, and is 
gravitationally bound, with the magnitude 
of its gravitational potential energy being greater than the sum of its rotational 
and thermal energies. Hence, we expect it to form a 
low mass star or substellar object depending on the efficiency with which it 
converts into a collapsed object. Figure 4 also shows a notable sharp drop in the disk
mass and an equivalent increase in the envelope mass at $t=0.26$~Myr. These effects are 
the manifestation of the ejection event when the clump leaves the parent disk and joins the envelope.
At about the same time, the stellar mass increases sharply due to consumption of the twin
clump that has lost its angular momentum due to the many-body interaction.

We have considered 20 models that exhibit disk fragmentation. The corresponding 
initial prestellar core masses lie in the range $M_{\rm c}=0.3-1.8~M_\odot$,
and ratios of rotational to gravitational energy are in the range 
$\beta=(0.25-2.3)\times 10^{-2}$. Only 11 models have shown ejection events 
and the parameters of these models are described in Table~\ref{table1}. 
A star symbol next to the model number means that the ejected clump
was initially a binary clump, and a dagger symbol means that the clump is smeared
out before reaching the computational boundary. Both these phenomena are discussed
later in this section.
The last five columns present the mass of the central star at the moment of ejection $M_{\rm s}$, 
the mass $M_{\rm eject}$ of the ejected clump, 
the time $t_{\rm eject}$ after the formation of the star when the ejection event takes place, 
the ejection velocity $v_{\rm
eject}$, and the escape velocity $v_{\rm esc}$, respectively. Both $v_{\rm eject}$ and $v_{\rm esc}$
are calculated at the time when the clump passes through the outer computational boundary. 
A diamond symbol indicates that the corresponding value is calculated at a smaller distance, 
since the ejected clump has been smeared out before reaching the outer computational boundary.
When more than one ejection event is present in the same model, the different
ejection speeds are separated by a slash.
In all cases, $v_{\rm eject}>v_{\rm esc}$, indicating that the ejected clumps will
be lost to the system and become freely-floating objects. 
The mean and median ejection velocities in our models are 0.7~km~s$^{-1}$ 
and 0.6~km~s$^{-1}$, and the minimum and
maximum values are 0.38~km~s$^{-1}$ and 1.2~km~s$^{-1}$, respectively.
These are consistent with
the velocities of low mass stars and BDs in the Chamaeleon~I and Taurus star-forming regions
\citep[e.g.,][]{Joergens06}. There is practically no
dependence of the ejection velocities on the mass of the ejected clumps.


Figure~\ref{fig5} presents the total mass in the system and illustrates the ejection events
in 11~models of Table~\ref{table1}. 
A sharp decrease in the total mass indicates the time when the clump escapes the computational region.
However, approximately half of the models (marked with a dagger symbol in Table~\ref{table1}) 
demonstrate a rather smooth decline in the total mass, extended over 
$(1-2)\times10^5$~yr. In these models, clumps have dispersed before reaching the outer computational
boundary. 
They may become gravitationally unbound due 
to strong tidal torques during the close encounters that lead to ejection, and/or due to 
decreasing numerical resolution on a logarithmically spaced grid as they propagate radially 
outward. 

Table~\ref{table1} demonstrates that some of the ejected clump masses $M_{\rm eject}$ 
are below the substellar mass limit,
giving us more confidence to propose this as a potential BD formation
mechanism. Note that because this process occurs due to the interactions of gaseous clumps and a gaseous disk, it is somewhat different from the scenario of ejections by 
multi-body interactions of point masses, i.e., objects that have already
fully collapsed down to compact equilibrium configurations.

Analysis of the ejection times in Table~\ref{table1} and Fig.~\ref{fig5} 
shows that most ejection events take
place between 0.2~Myr and 0.8~Myr after the formation of the central star 
and no ejections are seen after $t=1$~Myr. This is due to the fact that gravitational 
instability and fragmentation
are limited exclusively to the early stages of disk evolution.

It is interesting to note that models 6, 7, and 11 experience {\it multiple} 
ejection events. The same may
may be true for model~10 but the ejected masses at $t\approx0.25$ Myr and 
$t\approx0.4$ Myr are too small for the events to be resolved. In addition,
in models 2, 10, and 11 (marked with the star symbol), the ejected clumps were initially close 
binaries with separations of order a few AU, but later merged as they propagated toward the outer
boundary, most likely due to decreasing resolution at large radii. The possibility of ejection of 
close binary clumps is a particularly appealing feature of the clump ejection
scenario for BD formation since the
separation of binary BDs seems to peak at around $1-4$~AU \citep{Whitworth07}.

Fig.~\ref{fig6} shows the $\beta - M_{\rm c}$ phase space covered in our modeling. 
The dark-shaded 
area highlights the region where both disk fragmentation and clump ejection 
are observed, while the light-shaded
area defines the region where only disk fragmentation takes place.
There is a increasing tendency for ejections to take place as 
$\beta$ and $M_{\rm c}$ increase. 
Models that fall into the light-shaded domain usually form
just one fragment at a time, in which case the ejection mechanism cannot
work even in principle.
Furthermore, not all models that are in the ejection domain may undergo
ejection events due to the highly  stochastic nature of this mechanism. 
This means that the dark-shaded area outlines the region
where ejection is possible {\it in principle}, but in reality it might not be observed if
a chance arrangement of clumps would not favour their close encounters.
The robustness of Figure~\ref{fig6} is also affected by a rather coarse grid of models 
explored in the paper. For instance, we have considered only four values of $\beta$
and about ten models for each $\beta$. The uncertainty 
associated with a small number of models is illustrated in Figure~\ref{fig6} by arrows
showing where the fragmentation (and ejection) regions could potentially extend, 
had we considered a finer grid of models. 
We restricted our modeling with $M_{\rm c}\le2.0~M_\odot$ and ratios $\beta\le0.025$ 
due to the fact that models with higher initial core masses and angular momenta 
tend to form wide-separation ($> 10$~AU) binary/multiple stellar systems for which our 
numerical code is not well suited. Disks around such systems may
be strongly truncated due to tidal effects, which may impede disk fragmentation and ejection.
On the other hand, close encounters with other members of a stellar cluster 
may promote disk fragmentation and ejection \citep{Thies10}.
A dedicated study using suitable numerical codes is needed to address these issues.
We also note that a steep cutoff of the fragmentation region
at small $\beta\la 0.001$ is a physical effect related to the fact that such models form disks
of too low a mass and size for gravitational fragmentation to take place \citep{Vorob11}.


\bfig
  \resizebox{\hsize}{!}{\includegraphics{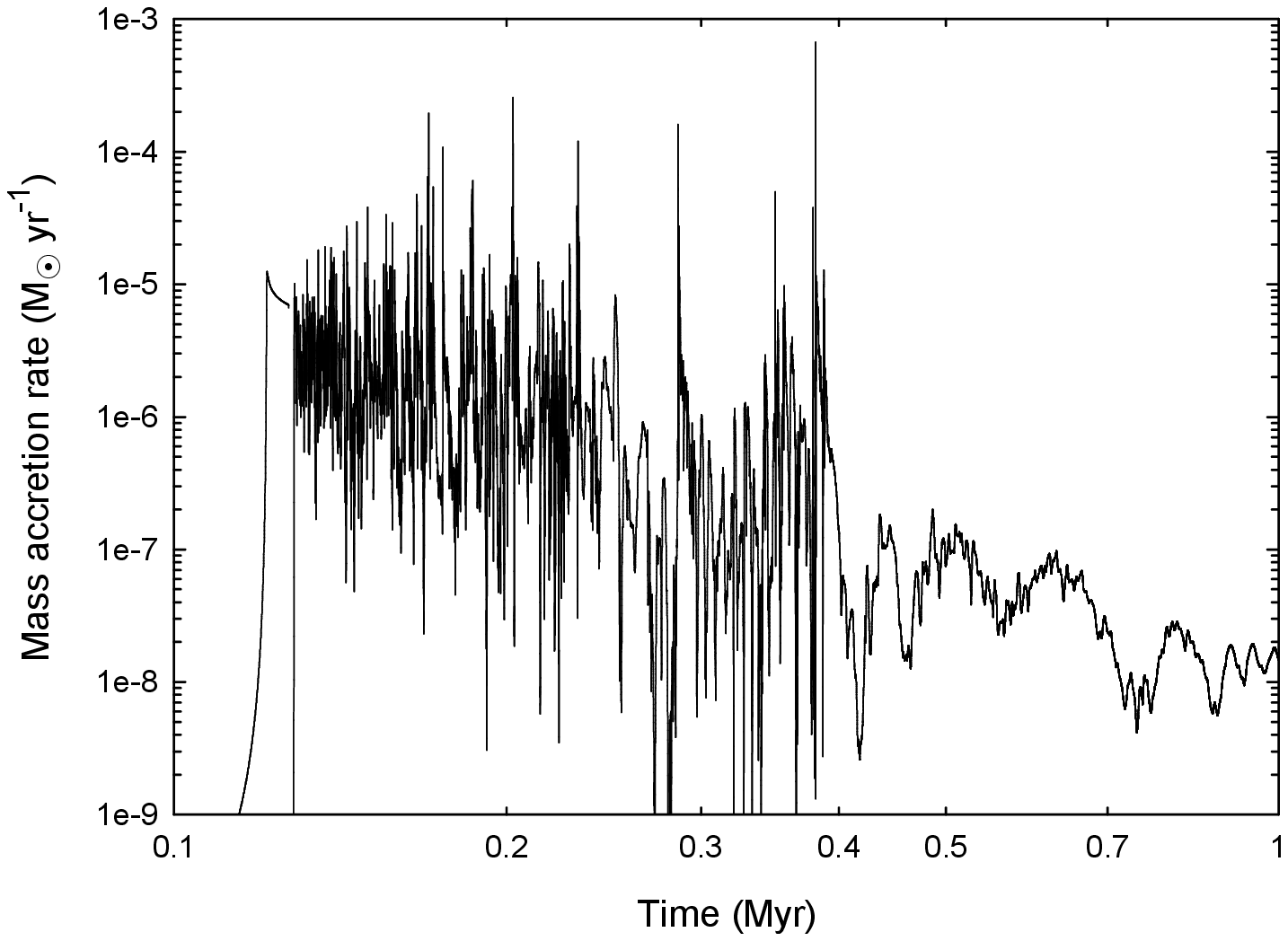}}
      \caption{Mass accretion rate onto the star as a function of the time elapsed
since the beginning of the collapse in the reference model. There is a particularly
large spike in mass accretion rate coinciding with the ejection of the clump
at 0.38 Myr (0.26~Myr after the formation of the central object).}
         \label{fig3}
\efig

\bfig
  \resizebox{\hsize}{!}{\includegraphics{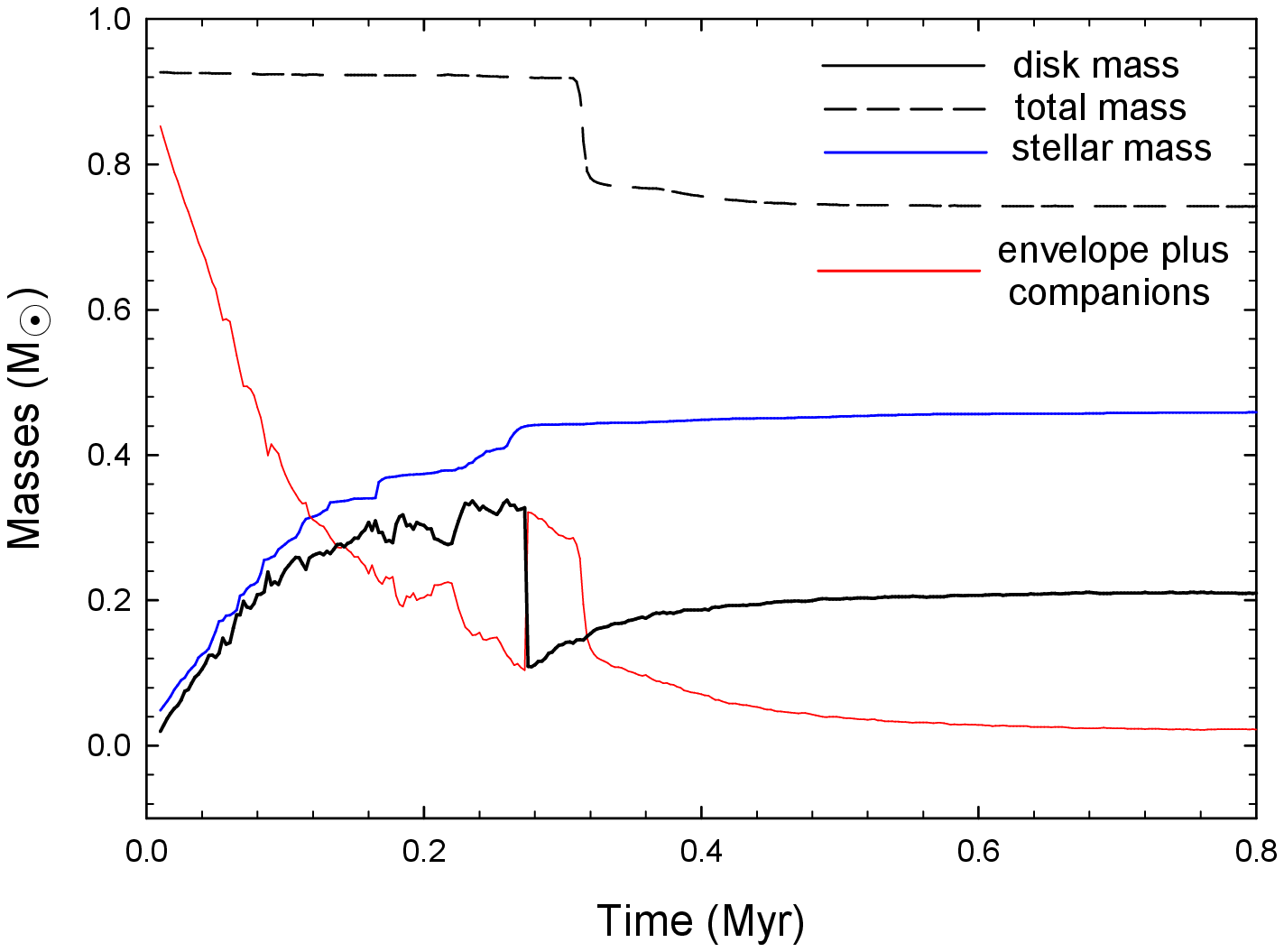}}
      \caption{Time evolution of the masses of the disk, star, envelope (including
ejected companions when present) and total mass as a function of time
since the formation of the star. Note the
large drop in disk mass and concurrent growth of envelope mass at about 
0.26 Myr, the time that the clump is ejected from the disk.}
         \label{fig4}
\efig

\bfig
  \resizebox{\hsize}{!}{\includegraphics{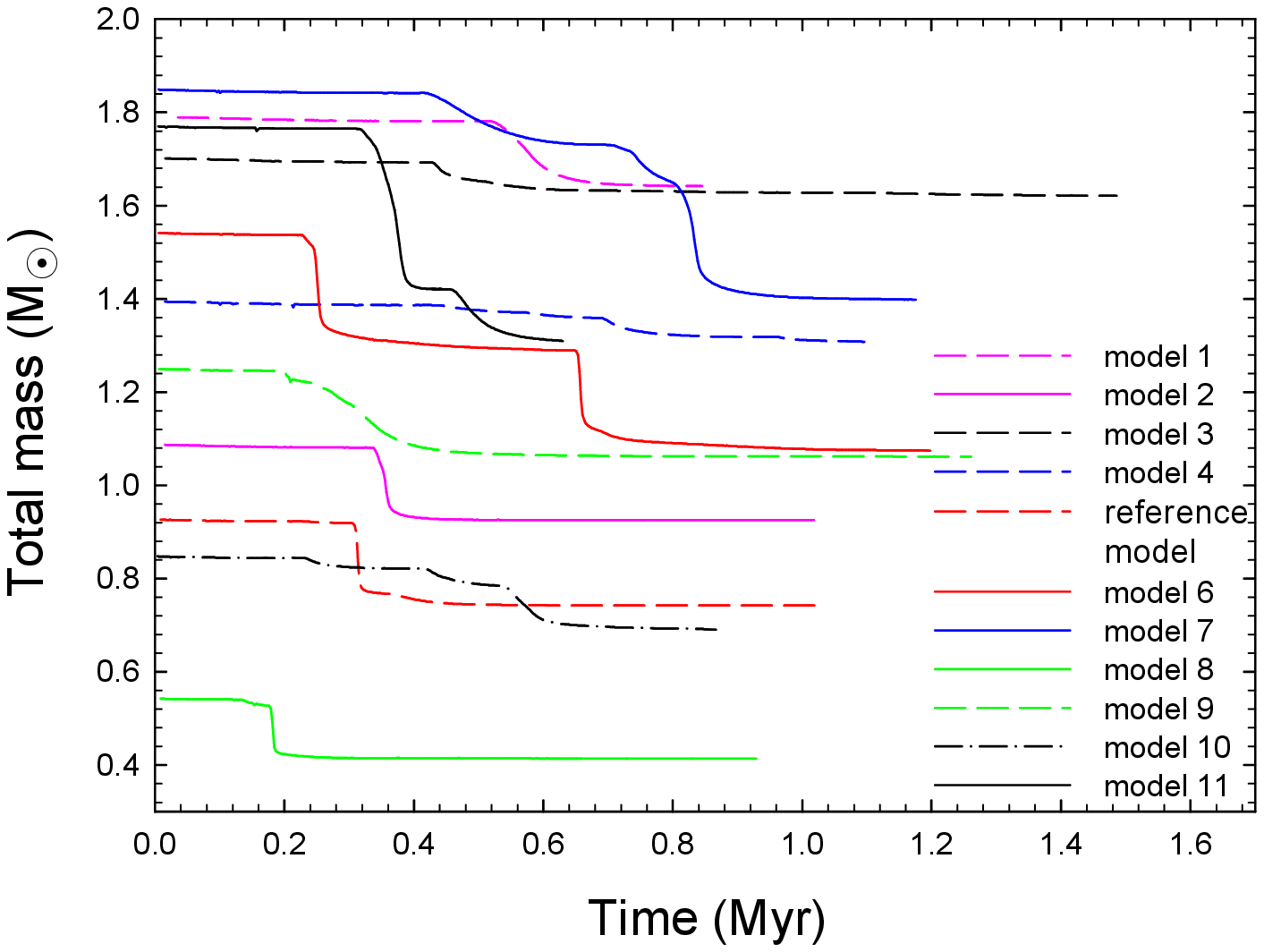}}
      \caption{Evolution of the total mass as a function of time since the formation of the star. 
      Note the large drop in mass at the time that a clump is ejected from the disk.}
         \label{fig5}
\efig

\bfig
  \resizebox{\hsize}{!}{\includegraphics{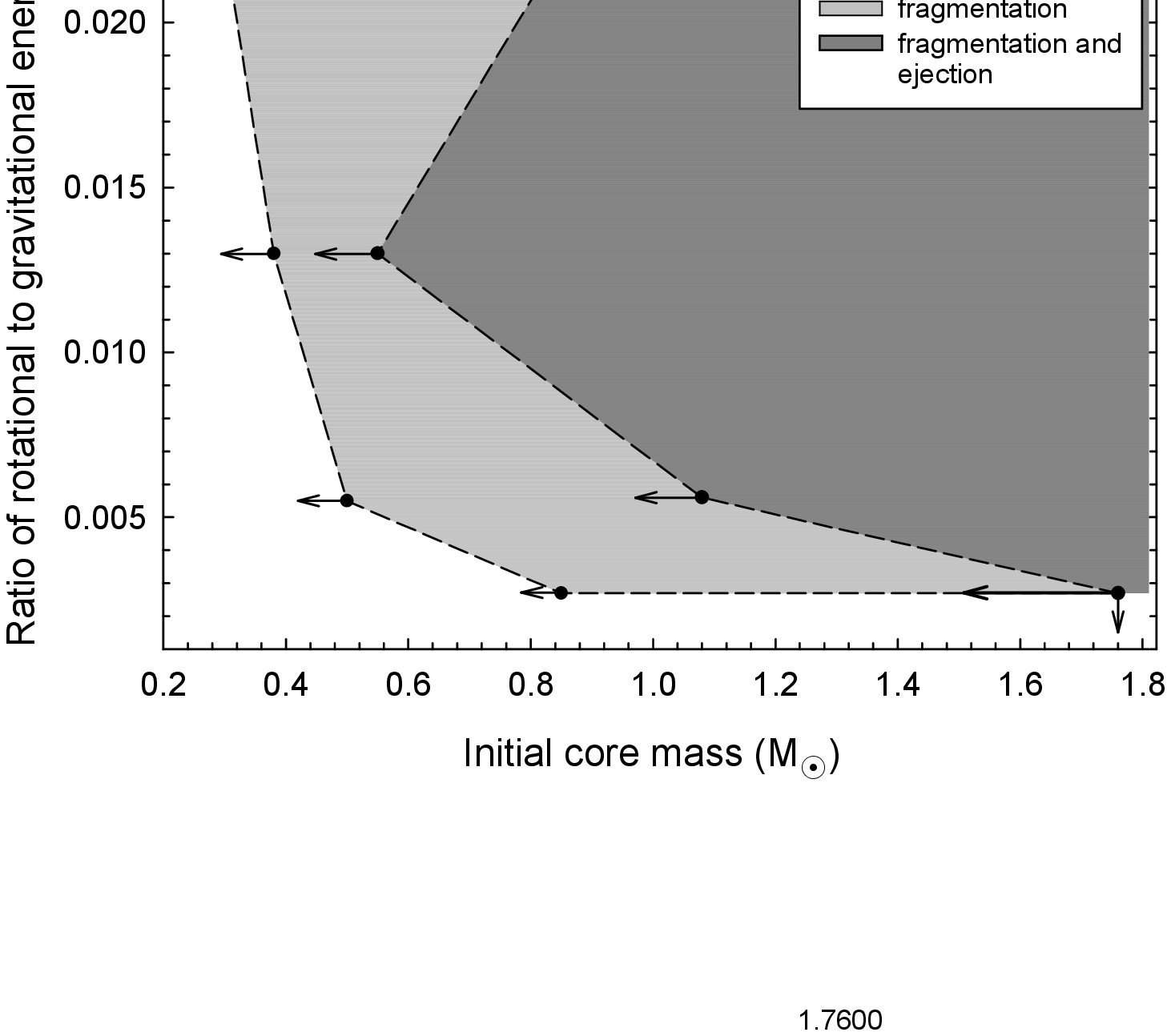}}
      \caption{Phase space of $\beta$, the initial ratio of rotational
to gravitational energy, versus core mass $M_{\rm c}$.
The solid dots correspond to various models in our study, the dark-shaded
region is the presumed region in which both fragmentation and ejection 
events may occur, and the light-shaded region is the one where only fragmentation events occur.
Arrows illustrate uncertainties associated with a coarse grid of models considered in the present study.}
         \label{fig6}
\efig

\section{Discussion and Conclusions}
\label{conclusions}

We find that high resolution models with a realistic treatment of heating
and cooling and
in which a disk forms consistently from prestellar collapse, can sometimes lead to
the ejection of gaseous clumps from the disk and eventually from the system 
(defined by the extent of the original prestellar core).
These are the first confirmed examples of ejections due to interactions of
compact gaseous clumps and their parent gas disks and host star, as opposed
to being caused by the interaction of point particles. 
Previous numerical simulations of ejections from disks by \citet{bat09} and 
\citet{sta09} have used sink particles with a smoothing length as a free
parameter. 
While the smoothing length can in principle be used to mimic the size 
of a clump, our
numerical simulations are free from such uncertainty and 
the size of the clump is instead determined self-consistently by 
the disk thermodynamics. We show explicitly that the clumps 
formed via disk fragmentation need {\it not} contract to stellar densities 
in order to be ejected into the intracluster medium. Moreover, some of 
the clumps may be tidally disrupted
during the ejection event and disperse as they leave the system, 
a phenomenon that is impossible to capture using sink particles. 
In fact, most of the clumps in the disk never achieve temperatures
required for a second collapse before they are disrupted, as pointed out 
previously \citep{Nayakshin10,bol10,vor11}. For instance, figure~2 (bottom-left panel) in \citet{vor11}
shows that most of the clumps have maximum temperatures below 1000~K and only one clump
achieves a temperature higher than 2000~K before being disrupted or torqued onto the star.
This makes the phase of
interaction of multiple first cores, as studied in this paper, 
all the more relevant to disk evolution.

An important argument usually in favor of the direct collapse scenario 
is the observation of some very low mass isolated proto-BD clumps
\citep{luh07}.
However, by allowing for the possibility of ejected clumps rather than 
finished BDs, our hybrid scenario also allows for this possibility.  
Clumps that survive can cool and contract
to form a (sub-)stellar core and a disk, thus also explaining the presence 
of accretion disks around BDs. 
Furthermore, the ejection speeds of the clumps are $ < 1$ km s$^{-1}
\approx 1$ pc Myr$^{-1}$, 
reflecting the fact that ejections take place in a much more diffuse
situation than interacting point masses in close proximity,
i.e., with impact parameters of tens of AU rather than several radii of 
finished BDs or low mass stars. With these
ejection speeds, the velocity dispersion of BDs is not expected to differ
much from that of YSOs and they should remain spatially co-located within 
cluster forming clumps of pc scale for at least the typical 1--2 Myr age of
YSO clusters.
Our models can also explain qualitatively why the BDs are less numerous than
stars, since the ejections only occur for some region of parameter
space. While a quantitative
prediction of BD fraction remains difficult without more insight
into core initial conditions, we can use a stellar IMF to obtain some estimates.
Assuming that about half the mass of each clump accretes onto a star
(i.e., 50\% formation efficiency), that only clumps with mass $M_{\rm c} \geq 0.8~M_\odot$
undergo ejection events, and using the Kroupa stellar IMF \citep{Kroupa01}
with lower and upper cutoff masses $0.08~M_\odot$ and $100~M_\odot$, respectively,
we estimate that the number of ejected clumps is about one for every 10 stars. 
Here we take into account that clumps survive ejection in only about half the realizations, 
but this decrease 
is mostly offset by models showing multiple ejection events. The net
ejection rate is then about one clump per model.   
This estimate is consistent with the estimated empirical ratio of 5--10 between stars and BDs \citep{luh07}.
We must keep in mind that our estimate is rather tentative and we expect it to increase
somewhat if more models are run at higher resolution.
There are also hints in our simulations that
ejected clumps with substantial rotation may form close binary objects,
in agreement with the prevalence of close binaries in BD-BD systems.
All in all, our hybrid scenario agrees, at least qualitatively, 
with nearly all of the key observational 
features \citep[see][]{luh07} of BDs. The only observation not easily 
accounted for are the wide BD binaries, but these are very rare and they
may indeed form from direct collapse since the total mass required is 
likely not substellar.

Our results reveal a fundamental process that is intrinsic to gaseous
disks with moderate to high levels of angular momentum. 
This clump ejection mechanism should be robust to environmental
variations of metallicity and temperature, as it depends more directly
on angular momentum and multi-body dynamics. Therefore, we believe that low mass stars
and substellar objects should form in a variety of situations in different
cosmic locales and epochs.

\acknowledgements
The authors thank the anonymous referee, as well as Pavel Kroupa and 
Alexander DeSouza, for helpful comments on the manuscript.
EIV gratefully acknowledges support 
from the RFBR grants 10-02-00278 and 11-02-92601 and also from a Lise Meitner Fellowship (FWF, Austria).
SB was supported by a grant from NSERC. We thank the Atlantic Computational 
Excellence Network (ACEnet) and the SHARCNET consortium for access to computational facilities.

\end{document}